\documentclass[doublecol]{epl2} %for 2 columns style without line numbers
\usepackage{amsmath}

\title{Star shaped patterns caused by colloidal aggregation during the spreading process of a droplet}
\shorttitle{Star shaped patterns caused by colloidal aggregation during the spreading process of a droplet} %Insert here a short version of the title if it exceeds 70 characters

\author{Michiko Shimokawa\inst{1} \and Hiroyuki Kitahata\inst{2} \and Hidetsugu Sakaguchi\inst{3}}
\shortauthor{M. Shimokawa \etal}

\institute{                    
  \inst{1} Fukuoka Institute of Technology - Wajiro-higashi, Higashi-ku, Fukuoka 811-0295, Japan\\
  \inst{2} Chiba University - Yayoi-cho 1-33, Inage-ku, Chiba 263-8522, Japan\\
  \inst{3} Kyushu University - Kasuga, Fukuoka 816-8580, Japan
}
\pacs{89.75.Kd}{Pattern formation in complex systems}
\pacs{05.65.+b}{Self-organized systems}
\pacs{47.20.Ma}{Interfacial instabilities}

\abstract{
This research found that when an acidic solution with a low surface tension 
spread on the surface of a glycerol solution mixed with milk, 
a star shaped pattern was spontaneously formed on the surface in the horizontal plane during the spreading process. 
We investigated the emergence of the star shaped pattern owing to an interfacial instability in experiments using glycerol solutions with several viscosities and 2-methoxyethanol aqueous solutions, which are acidic solutions, with several concentrations.
This result demonstrated that the star shaped pattern emerged in the high concentration of 2-methoxyethanol. 
We proposed a phenomenological model, based on our experimental results, which explains three points as follows; 
the spreading of the 2-methoxyethanol aqueous solution on the surface of the glycerol solution, 
the colloidal aggregation of the milk protein colloids 
caused by the denaturation that occurs when mixed with 2-methoxyethanol, 
and the accumulation of the 
aggregates toward the dent regions of the moving interface by a sweeping effect. 
The model reproduces the formation of the star shaped pattern which was similar to the experimental one.
Furthermore, the model provided a phase diagram against the concentration of the 2-methoxyethanol solution and the viscosity of the glycerol solution as control parameters in our experiments.
The phase diagram was close to that obtained from our experiments. 
The results suggest that the above three points are important for the formation of the star shaped pattern.}

\begin{document}

\maketitle
\section{Introduction}
Patterns are abundant in nature~\cite{pattern1,pattern2,pattern3,pattern4,pattern5,river1,river2,columnar1,columnar2,columnar3,bacteria1},
as seen in snowflakes~\cite{pattern5}, the branching of rivers~\cite{river1,river2}, columnar joints~\cite{columnar1,columnar2,columnar3},
and bacteria colonies~\cite{bacteria1}.
Instabilities at interfaces play important roles for most of these pattern formations.
Typical interfacial instabilities found within fluid systems 
include the Rayleigh-Taylor instability, Saffman-Taylor instability and Kelvin-Helmholtz instability~\cite{instability1,instability2,VF,ST1,KH1,KH2},
which occur at the interface of two solutions due to density, 
viscosity and velocity differences.

Interfacial instabilities have also been studied in polymer chemistry~\cite{gels1,gels2}, colloid science~\cite{Ring1,Ring2,Ring3,Fractal1,Cell,colloid1,colloid2,colloid3}, granular physics~\cite{Grain1,Grain2, Grain3, Grain4} and crystal growth~\cite{MS1,MS2}.
In a polymer system such as an elastic material, a meandering pattern is formed due to the instability of a crack front~\cite{gels1,gels2}. 
In a colloidal system, particles with a colloidal size aggregate and 
the instability due to the aggregations leads to formations of multiple patterns~\cite{Ring1,Ring2,Ring3,Fractal1,Cell,colloid1,colloid2,colloid3}.
The Rayleigh-Taylor instability and Saffman-Taylor instability emerge even in granular systems~\cite{Grain1,Grain2, Grain3, Grain4}.
In crystal growth, a flat interface becomes unstable by the Mullins-Sekerka instability caused by competitive effects between the diffusion field of concentration or temperature and the surface tension, and as a result, a dendritic shape appears~\cite{MS1,MS2}.

In the Saffman-Taylor instability and Mullins-Sekerka instability, protruded regions or tip regions in the interface grow faster, and the tip structure becomes sharper. The diffusion-limited aggregation (DLA) is an ideal model of this type of interfacial instability~\cite{DLA1,DLA2,DLA3}.  In this paper, we show a different type of interfacial instability. 
In the instability, the dent regions in the interface grow slower than the surroundings
due to the interference of the colloidal aggregates~\cite{gelation1,gelation2}, 
which generates a star shaped pattern 
but does not lead to a more complicated pattern 
such as a DLA-like pattern via successive tip-splittings~\cite{split1,split2}.

In the experiment, an acidic solution with a low surface tension spread on a surface of a glycerol solution, and mixed with milk.
Then, milk protein colloids were aggregated by 
the denaturation that occurs when mixed with the acidic solution. 
The aggregates cause the interfacial instability between two solutions, 
leading to the formation of a star shaped pattern.
In Secs.~II and III, we demonstrate the experimental results of the star shaped pattern,  and discuss the pattern formation through a phenomenological model in Sec.IV.

\section{Experimental Methods}
Either magnetic fluid (TaihoKozai W-40)
with $\rho_{d1}=1.4$ g/ml
or 2-methoxyethanol aqueous solution
with $0.97$ g/ml$\leq\rho_{d2}\leq0.99$ g/ml
was prepared to drop onto a glycerol aqueous solution
with $1.1$ g/ml $\leq\rho_{g}\leq1.2$ g/ml.
2-methoxyethanol aqueous solution was prepared using 2-methoxyethanol (Wako 055-01096) 
with acidity and water with a concentration $c$,
which is defined as a volume ratio of 2-methoxtethanol to water.
Food colorings (Kyoritsu Syokuhin) with red and green colors 
were mixed into the 2-methoxyethanol solution for visualization of the surface pattern~
\cite{food}.
The viscosity of the 2-methoxyethanol solution almost kept constant against $c$. 
The glycerol solution was prepared using glycerol (WAKO 072-00621), milk (Aso-Kogen Tokusen Gyunyu, Glico), and water. 
Milk was mixed with the mixed solution of glycerol and water at a volume ratio 1:20.
The viscosity $\mu$ of the glycerol solution increases with the amount of glycerol. 
A temperature of 25$^\circ$C was maintained in our experiments, as viscosity is sensitive to temperature. 

\begin{figure}[h]
	\begin{center}
		\includegraphics[width=4.5cm,clip]{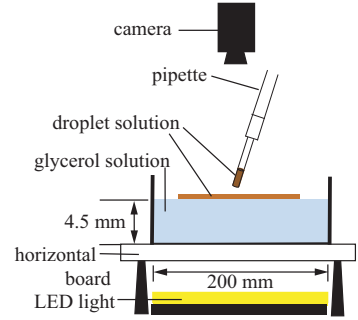}
	\end{center}
\caption{Experimental setup.}
\label{method}
\end{figure}

The experimental procedure was as follows: 
A square container with 200 mm as side lengths was placed on a clear horizontal board.
A LED light was put under the clear horizontal board for visualizations in experiments with the 2-methoxyethanol aqueous solution. 
Glycerol solution was poured into the container until it reached the 4.5 mm depth, as shown in Fig.~\ref{method}. 
The container was left to stand for 10 min or more to allow the fluid to settle. 
Next, a droplet of magnetic fluid or 2-methoxyethanol aqueous solution with the volume of 0.01 ml was carefully placed at the center of the surface of the glycerol solution with a micropipette (Gilson Pipetman). 
As the solution of a droplet has a low surface tension 
and the interfacial tension between two solutions does not exist because of the miscible behavior, 
the solution of the droplet spreads on the surface of the glycerol solution  
due to Marangoni effect~\cite{marangoni,marangoni1}.
In other words, $\sigma_{d1}<\sigma_{g}$ and $\sigma_{d2}<\sigma_g$ lead to the spreading of the droplet on the surface of the glycerol solution, where $\sigma_{d1}\sim28$ mN/m, $\sigma_{d2}\sim30.5$ mN/m and $\sigma_{g}\sim49.1$ mN/m represent the surface tensions of the magnetic solution, the 2-methoxyethanol aqueous solution, and glycerol solution, respectively.
The surface pattern was recorded by a digital camera (Canon EOS kiss X6i) from a top view of the container.

\section{Experimental results}
\subsection{Star shaped pattern by magnetic fluid}
\begin{figure}[h]
	\begin{center}
		\includegraphics[width=8.5cm,clip]{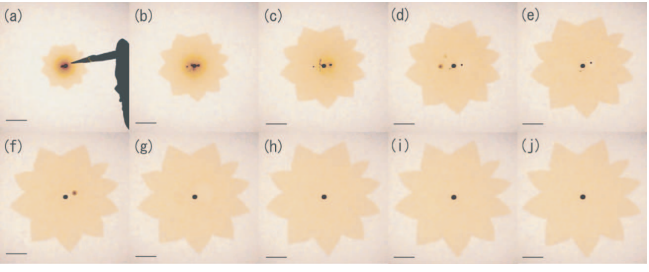}
	\end{center}
	\caption{Snapshots of the star shaped pattern observed in experiments with magnetic fluid at time (a) $t=$ 0.5 s, (b) 1.0 s, (c) 1.5 s, (d) 2.0 s, (e) 2.5 s, (f) 3.0 s, (g) 3.5 s, (h) 4.0 s, (i) 4.5 s, and (j) 5.0 s, where $t=0$ is the time when the droplet of magnetic fluid was put on the surface of glycerol solution with the viscosity of 7.76 mPa$\cdot$s. Scale bar: 30 mm.}
\label{magnetic}
\end{figure}

In this section, we show the experimental results in a use of magnetic fluid as a droplet.
Figure~\ref{magnetic} shows snapshots of the surface pattern drawn by magnetic fluid, 
captured from above as shown in Fig.~\ref{method}. 
As soon as the droplet of magnetic fluid was put on the surface of the glycerol solution,
magnetic fluid spread on the surface without sinking,
i.e., the time scale of the spreading is sufficiently shorter than that of the sinking
even if $\rho_{d1}>\rho_{g}$,
and it is considered that the vertical behavior is not effective for the formation of the surface pattern
~\cite{Fractal1,Cell}.
Then, the interface between the magnetic fluid and glycerol solution deformed to a star shaped pattern as shown in Fig.~\ref{magnetic}.
The star shaped pattern grew with the similar number of 
protrusions.
The interface growth occurred until $t\sim3$ s (Figs.~\ref{magnetic}(a)--(e)), 
where $t=0$ was the time when the droplet was put on the surface of glycerol solution.
After the growth stopped around at $t\sim3$ s (Fig.~\ref{magnetic}(f)), 
the star shaped pattern was kept without the shrinkage (Figs.~\ref{magnetic}(f)--(j)).

We prodded the interface between two solutions with a pick, 
and colloidal sheets were formed in the dent regions of the star shaped pattern.
It is considered that the formation of the colloidal sheet at the interface of two solutions is owing to colloidal aggregations of milk protein colloids, included in the glycerol aqueous solution, due to the denaturation that occurs when mixed with the acidic solution included in the magnetic fluid~\cite{gelation1, gelation2}.
On the other hand, the sheets were not observed in the protruded regions.

\subsection{Star shaped pattern by 2-methoxyethanol aqueous solution}
The major components of magnetic fluid are magnetite particles, 
surfactant (alcohol) and water (or oil).
2-methoxyethanol, whose aqueous solution is acidic, was used as a surfactant for the magnetic fluid we used.
We investigated the surface pattern in the experiments with 2-methoxyethanol aqueous solution in the place of magnetic fluid
in order to understand the formation process of the star shaped pattern deeply by a simple experimental setup.

\begin{figure}
	\begin{center}
		\includegraphics[width=8.5cm,clip]{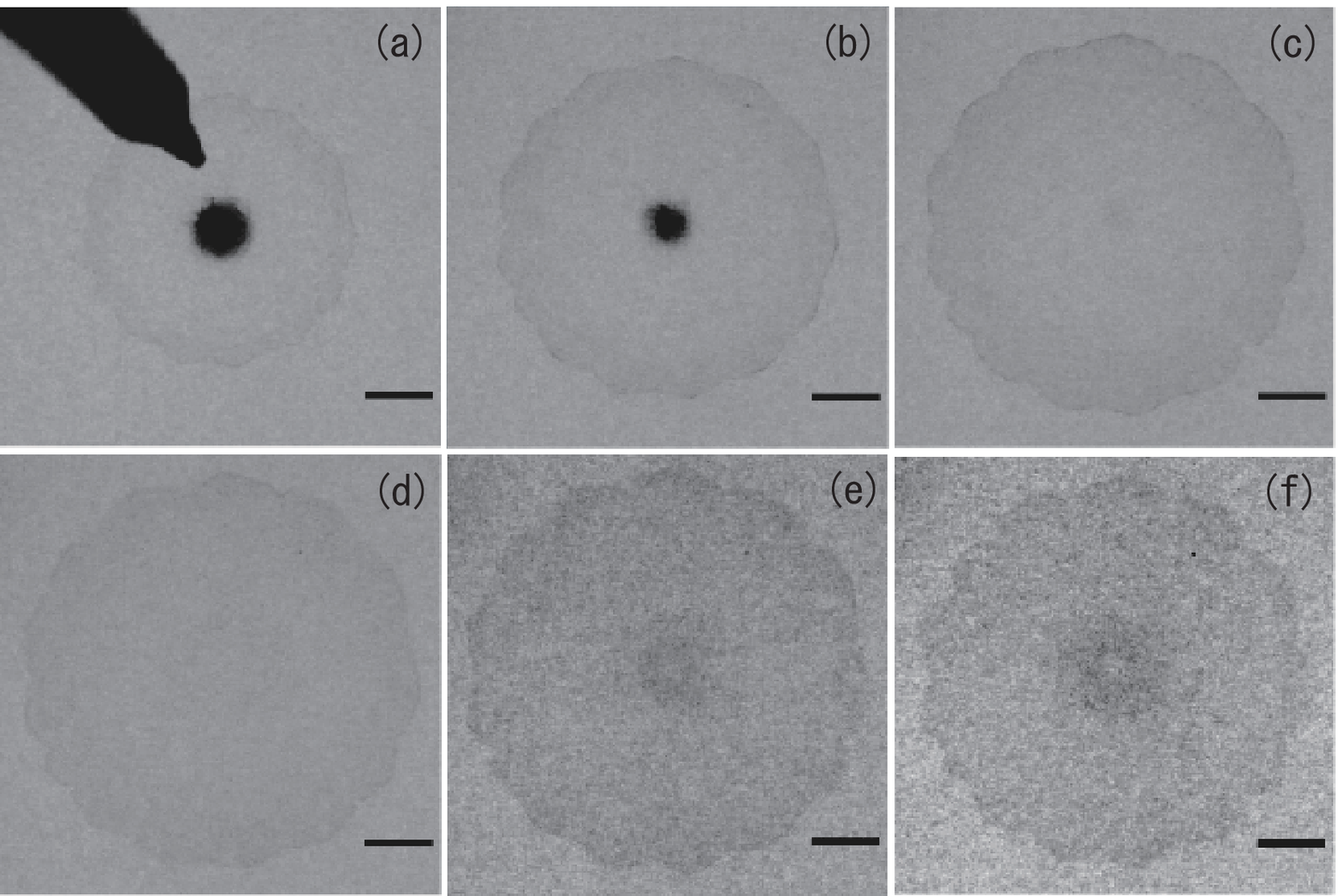}
	\end{center}
\caption{Snapshots of the star shaped pattern observed in a spreading process at (a) $t=$ 0.5 s, (b) 1.0 s, (c) 1.5 s, (d) 2.0 s, (e) 2.5 s, and (f) 3.0 s,
where $t$ = 0 corresponds to the time when the droplet of 2-metoxyethanol aqueous solution was put on the surface of the glycerol solution. 
		These snapshots were observed in a use of the glycerol solution with the viscosity of $\mu =30$ mPa$\cdot$s and 2-methoxyethanol aqueous solution with the concentration $c=100\%$ colored by adding food color. Scale bar: 10 mm.}
	\label{timestar}
\end{figure}

Figure~\ref{timestar} shows the snapshots of the surface pattern, captured from above as shown in Fig.~\ref{method}. 
The patterns were observed at (a) $t=0.5$ s, (b) 1.0 s, (c) 1.5 s, (d) 2.0 s, (e) 2.5 s, and (f) 3.0 s,
where $t=0$ was the time when a droplet of 2-methoxyethanol solution with $c=100\%$ was put on the surface of the glycerol solution with a viscosity $\mu=30$ mPa$\cdot$s.
Since the surface tension of the 2-methoxyethanol aqueous solution was sufficiently lower than that of the glycerol solution,
the 2-methoxyethanol solution spread on the surface of the glycerol solution soon
after the droplet was put on the surface of glycerol solution (Figs.~\ref{timestar}(a)--(c)).
In the spreading process, the star shaped pattern 
was formed spontaneously.
The average value of the spatial growth rate decreased with time, and the growth stopped (Figs.~\ref{timestar}(d)--(f)). 
The number of protrusions
of the star shaped pattern were kept constant in the spreading process.
Not only the pattern but also the dynamics before $t\sim3$ s was similar to that in Fig.~\ref{magnetic}.
The results imply that the effect caused by 2-methoxyethanol was important for 
the initial stage of the formation of the star shaped pattern.

\begin{figure}
	\begin{center}
		\includegraphics[width=8.5cm,clip]{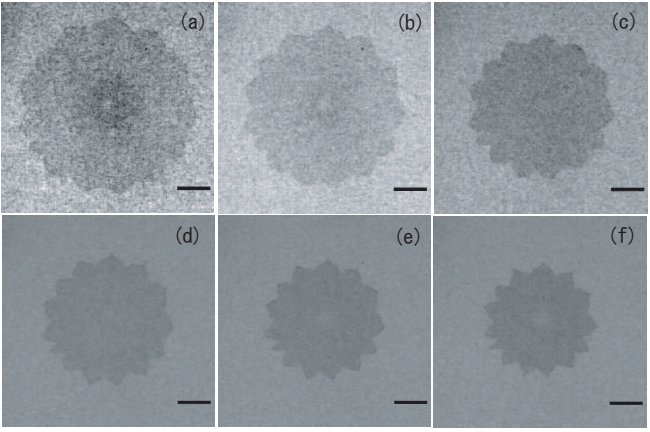}
	\end{center}
\caption{Snapshots of the star shaped pattern captured in the shrinking process at time (a) $t=$ 3.5 s, (b) 4.0 s, (c) 4.5 s, (d) 5.0 s, (e) 5.5 s, and (f) 6.0 s,
where $t$ = 0 corresponds to the time when the droplet of 2-methoxyethanol aqueous solution was put on the surface of the glycerol solution. 
These snapshots were observed in a use of the glycerol solution with the viscosity of $\mu =30$ mPa$\cdot$s and the 2-methoxyethanol solution with the concentration of $c=100\%$. Scale bar: 10 mm.}
	\label{timestar1}
\end{figure}

As shown above, the behavior in the spreading process was similar to that in the experiments with magnetic fluid.
The behavior after the spreading process, however, was different from that in a use of magnetic fluid.
After $t=3.5$ s, the pattern shrank as shown in Fig.~\ref{timestar1}.
The star shaped pattern also had the similar number of protrusions
in the shrinking process,
although the amplitude of the protrusions increased with time (Figs.~\ref{timestar1}(a)--(d)). 

We also observed the shrinking process in a use of 
ethanol solution
in the place of 2-methoxyethanol aqueous solution,
although the star shaped pattern was not formed.
This result provides that the shrinking process is 
owing to the vaporization of the component of the droplet,
as 2-methoxyethanol and ethanol are both volatile liquids.
It is considered that the shrinking 
process is not important for the interfacial instability which leads to a formation of the star shaped pattern
since the protrusions emerge before the shrinking process as shown in Fig.~\ref{timestar}. 

\begin{figure}[h]
\begin{center}
\includegraphics[width=8.5cm,clip]{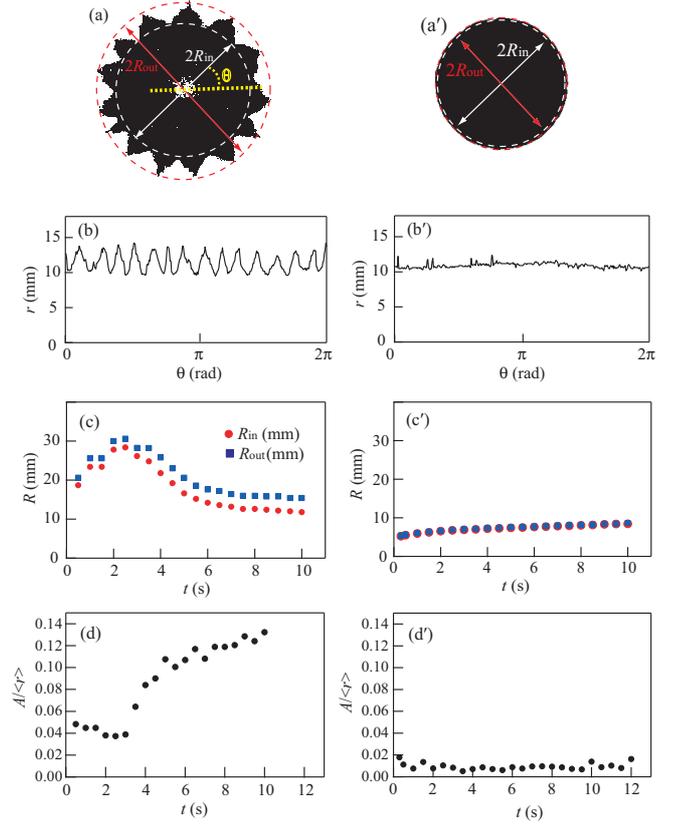}
\end{center}
\caption{(a)--(d) Star shaped pattern for $c=100\%$ and (a')--(d') circular pattern for $c=20\%$, 
where $c$ is the concentration of 2-methoxyethanol solution.
The viscosity of the glycerol solution was fixed at $\mu=30$ mPa$\cdot$s.
(a), (a') Binary images of 
surface patterns.
(b), (b') Distances $r$ from the center of mass of the pattern
to the edge of pattern for (a) and (a') against an azimuthal direction $\theta$, respectively.
(c), (c') Inner circle radius $R_{\text{in}}$ and outer circle radius $R_{\text{out}}$ for patterns of (a) and (a') at $t$, where $t=0$ is the time that the droplet of 2-methoxyethanol aqueous solution is put on the surface of the glycerol solution.
(d), (d') Time series of $A/\left< r \right>$ for patterns of (a) and (a'), 
where $A=(R_{\text{out}}-R_{\text{in}})/2$ and $\left< r \right>=(R_{\text{out}}+R_{\text{in}})/2$.}
\label{expdata}
\end{figure}

Next, we investigated the surface pattern in a use of 2-methoxyethanol aqueous solution with several concentrations $c$.
The star shaped pattern in Fig.~\ref{expdata}(a) transformed to a circular pattern in Fig.~\ref{expdata}(a') with a decrease in $c$.
These patterns, observed in $\mu=30$ mPa$\cdot$s, emerged at $c=100\%$ and $20\%$, respectively.
We measured quantitative data for these patterns, 
where the data of Figs.~\ref{expdata}(b)-(d) and Figs.~\ref{expdata}(b')-(d') were obtained from the videos
of the formations of the star shaped pattern and the circular pattern, respectively.
Figures~\ref{expdata}(b) and (b') show the distance $r$ from the center of mass of the pattern to the interface against an azimuthal direction $\theta$ as shown in Fig.~\ref{expdata}(a).
The results demonstrate that the radius of the star shaped pattern was close to that of the circular pattern,
and that the amplitude $A$ of the star shaped pattern was sufficiently larger than that of the circular pattern.
The number of protrusions was 14 for the pattern in Fig.~\ref{expdata}(a), 
and was kept constant in a time development as shown in Figs.~\ref{timestar} and \ref{timestar1}.
Next, we measured an inner radius $R_{\text{in}}$ and an outer radius $R_{\text{out}}$, 
as shown in Figs.~\ref{expdata}(a) and (a').
Figures~\ref{expdata}(c) and (c') are $R_{\text{in}}$ and $R_{\text{out}}$ at $t$, 
where $t=0$ is the time when the droplet of 2-methoxyethanol aqueous solution was put on the surface of glycerol solution.
In the formation of the star shaped pattern, $R_{\text{in}}$ and $R_{\text{out}}$ increased with time, 
and started decreasing at around 
$t\sim2.5$ s (Fig.~\ref{expdata}(c)).
On the other hand, in the formation of the circular pattern, $R_{\text{in}}$ and $R_{\text{out}}$ increased with time, 
and were saturated $R_{\text{in}}\sim$$R_{\text{out}}\sim$10 mm (Fig.~\ref{expdata}(c')).
It is considered that a shrinkage of the pattern is derived from the vaporization of 2-methoxyethanol included in the droplet solution, 
as the value of $c$ in the experiment of Fig.~\ref{expdata}(c) is higher than that in the experiment of Fig.~\ref{expdata}(c').
This result supports our suggestion.

Next, we investigated a dimensionless parameter $A/\left< r \right>$ at $t$, 
where $A=(R_{\text{out}}-R_{\text{in}})/2$ and $\left< r \right>=(R_{\text{in}}+R_{\text{out}})/2$.
In the formation of the star shaped pattern (Fig.~\ref{expdata}(d)), 
$A/\left< r \right>$ started increasing around at $t\sim3.0$ s, 
and was saturated at $A/\left< r \right>\sim0.14$.
On the other hand, in the formation of the circular pattern (Fig.~\ref{expdata}(d')),
$A/\left< r \right>$ was kept almost constant at $A/\left< r \right>\sim0.01$ in the time development.
The value $A/\left< r \right>\sim0.01$ was sufficiently smaller than $A/\left< r \right>\sim0.14$ in the case of the star shaped pattern.
We considered $A/\left< r \right>$ as a characteristic parameter to determine whether the star shaped pattern emerged.

\begin{figure}[h]
\begin{center}
\includegraphics[width=8.5cm,clip]{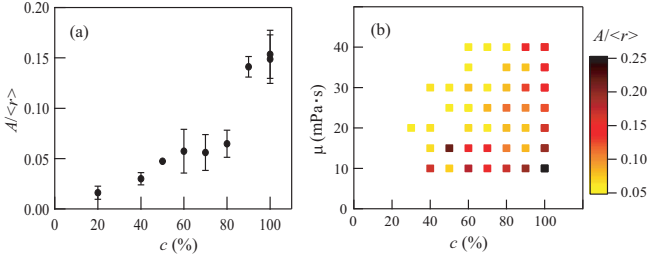}
\end{center}
\caption{(a) $A/\left< r \right>$ against 2-methoxyethanol aqueous solution with the concentration $c$ 
in experiments of the glycerol solution with the viscosity of $\mu=40$ mPa$\cdot$s, 
where $A=(R_{\text{out}}-R_{\text{in}})/2$ and $\left< r \right>=(R_{\text{out}}+R_{\text{in}})/2$.
(b) Phase diagram against $\mu$ and $c$. Color bar shows a value of $A/\left< r \right>$ for each patterns.}
\label{phase}
\end{figure}

Figure~\ref{phase}(a) shows $c$ dependence of $A/\left< r \right>$ in experiments with $\mu=40$ mPa$\cdot$s.
The $A/\left< r \right>$ jumps at $A/\left< r \right>\sim0.05$, 
where the circular shaped pattern transforms to the star shaped pattern.
By setting the threshold $A/\left< r \right> = 0.05$, 
we investigated a phase diagram against $c$ and $\mu$ in Fig.~\ref{phase}(b).
The color bar shows an intensity of $A/\left< r \right>$.
The result in Fig.~\ref{phase}(b) demonstrates 
that the star shaped pattern is formed for larger $c$.

\section{Phenomenological model}
\begin{figure}[h]
\begin{center}
\includegraphics[width=7cm,clip]{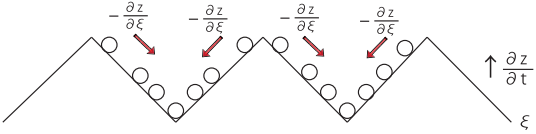}
\end{center}
\caption{Schematic to show the effect of the last term in Eq.~(\ref{model1-2}).}
\label{n1}
\end{figure}

We propose a simple phenomenological model to qualitatively understand the formation mechanism of the star shaped pattern.  
We assume that the following three points were important for the formation of the star shaped pattern:
(1) the spreading of the acidic solution with a low surface tension on the glycerol solution, 
caused by
the Marangoni effect,
(2) colloidal aggregation of milk protein colloids caused by the denaturation that occurs when mixed with the 2-methoxyethanol included in a droplet solution,
(3) the accumulations of colloidal aggregations
swept into the dent regions.
We focus on the dynamics about the density of colloidal aggregation $n$ and the distance of the growing interface from the origin $z$.
A simple one-dimensional interface growth model incorporating the three points is expressed as 
\begin{eqnarray}
	&&\frac{\partial z}{\partial t}=ac(1-n)+\alpha(n)D_z\frac{\partial^2 z}{\partial \xi^2}\label{model1-1},\\
	&&\frac{\partial n}{\partial t}=\beta\frac{\partial z}{\partial t}+D_n\frac{\partial^2 n}{\partial \xi^2}+\gamma\frac{\partial}{\partial \xi}\left(\frac{\partial z}{\partial t}\frac{\partial z}{\partial \xi}n\right)\label{model1-2},
\end{eqnarray}
where $\xi$ denotes the position along the interface, 
and $a$, $c$, and $\beta$ are positive control parameters.
For Eq.~(\ref{model1-1}), the first term $ac(1-n)$ implies the interface growth by the denaturation and the Marangoni effect.
The second term $\alpha(n)D_z\partial^2z/\partial \xi^2$ expresses the effect of the effective line tension, 
where $\alpha(n)$ is generally a function of $n$ but $\alpha(n)$ is set to be 1 in the simplest case. 
For Eq.~(\ref{model1-2}), the first term $\beta \partial z/\partial t$ implies 
that the denaturation of milk protein colloids occurs in proportion to the advance of the interface.
The second term $D_n\partial^2n/\partial \xi^2$ represents the diffusion of 
colloidal aggregates generated by the denaturation.
The last term $\gamma\partial/\partial \xi\{(\partial z/\partial t)(\partial z/\partial \xi) n\}$ represents the sweeping effect, that is, colloidal aggregates tend to be accumulated to the dent region as $-\partial J/\partial \xi$  where the flow $J$ is proportional to the interface velocity $\partial z/\partial t$, interface slope $-\partial z/\partial \xi$, and $n$ as shown in Fig.~\ref{n1}.

For a flat interface, $z(\xi,t)$ and $n(\xi,t)$ are independent of $\xi$, then, $z(\xi,t)=z_0(t)$ and $n(\xi,t)=n_0(t)$ satisfy
\begin{eqnarray}
&\dfrac{dz_0}{dt}=ac(1-n_0),\\
&\dfrac{dn_0}{dt}=\beta\dfrac{dz_0}{dt}=a \beta c(1-n_0).\
\label{model0}
\end{eqnarray}
By the linear stability analysis around this flat interface solution, 
$z(\xi,t)=z_0(t)+\sum_{k\neq0} z_k(t)e^{ik\xi}$ and $n(\xi,t)=n_0(t)+\sum_{k\neq0} n_k(t)e^{ik\xi}$,
where $z_k(t)e^{ik\xi}$ and $n_k(t)e^{ik\xi}$ are spatial perturbations, 
satisfy the following equations:
\begin{eqnarray}
&\dfrac{dz_k}{dt}=-acn_k-D_zk^2z_k,\\
&\dfrac{dn_k}{dt}=\beta\dfrac{dz_k}{dt}-D_nk^2n_k-\gamma k^2 ac(1-n_0)n_0z_k.\
\label{linearlity}
\end{eqnarray}

The linear growth rate $\lambda$ of the spatial perturbations with the wavenumber $k$ is determined by the equation
\begin{align}
\lambda^2&+(D_nk^2+D_zk^2+ac\beta)\lambda\nonumber\\
&+D_nD_zk^4-\gamma ac k^2n_0ac(1-n_0)=0.\
\label{solution}
\end{align}

\begin{figure}[h]
\begin{center}
		\includegraphics[width=8.5cm,clip]{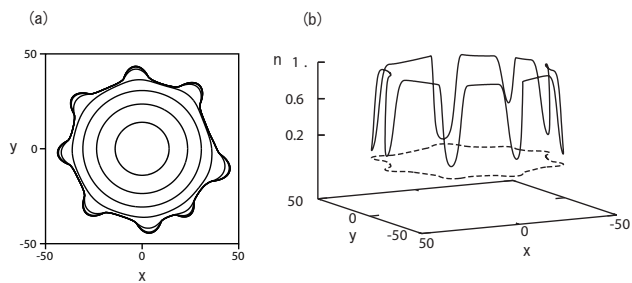}
	\end{center}
	\caption{(a) Snapshots of patterns obtained by Eqs.~(\ref{model1-1}) and (\ref{model1-2}) with $\alpha(n)=1-n$ at $a=0.01$, $c=1$, $D_z=0.04$, $\beta=0.02$, $D_n=0.04$ and $\gamma=10$ in the space of ($x$, $y$).
		(b) Three-dimensional plot of the concentration $n$ (solid line) of colloidal aggregation on the star shaped interface (dashed line).}
	\label{model2}
\end{figure}

For sufficiently small $k$, Eq.~(\ref{solution}) has the solution:  
\begin{eqnarray}
\lambda_k\simeq \dfrac{ac}{\beta}\gamma n_0(1-n_0)k^2.
\label{lambda}
\end{eqnarray}
For $\beta>0$, $\lambda_k$ is positive and therefore, the flat interface is unstable. 
This is because the 
colloidal aggregates tend to gather in the dent regions in the growing interface due to the sweeping effect, and the density increase in the dent regions slows down the interface growth, which makes the dents deeper, as shown in Fig.~\ref{model2}. 

We have performed numerical simulations of a radial growth of the interface using the one-dimensional model equation for the sake of simplicity. For the radial growth, the average value of $z$ is calculated first and is expressed as $R$. The position $\xi$ is interpreted as the coordinate along the azimuthal direction that $\xi=R\theta$. The one-dimensional equations (\ref{model1-1}) and (\ref{model1-2}) are calculated under the periodic boundary conditions $z(0)=z(2\pi R)$ and $n(0)=n(2\pi R)$.  

\begin{figure}[h]
	\begin{center}
	\includegraphics[width=8.5cm,clip]{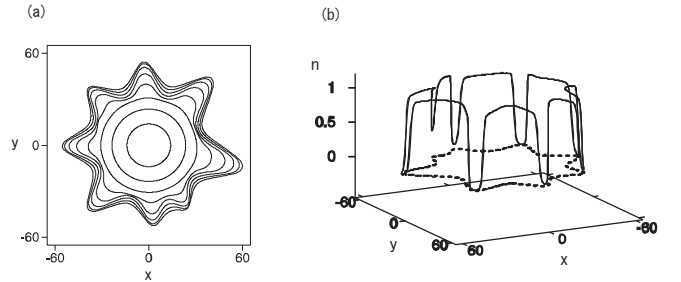}
	\end{center}
	\caption{(a) Star shaped pattern obtained by the numerical simulation of Eqs.~(\ref{model1-1}) and (\ref{model1-2}) with $\alpha(n)=1$ at $a=0.01$, $c=1$, 
		$D_z=0.02$, $\beta=0.02$, $D_n=0.03$, and $\gamma=10$ in the space of ($x$, $y$).
		Solid lines represent the snapshots of interface patterns at different times. 
		(b) Three-dimensional plot of the concentration $n$ (solid line) of colloidal aggregation on the star-shaped interface (dashed line).}
	\label{model1}
\end{figure}

Figure~\ref{model1} was obtained by numerical simulation of Eqs.~(\ref{model1-1}) and (\ref{model1-2}) at $a=0.01$, $c=1$, $D_z=0.02$, $\beta=0.02$, $D_n=0.03$, and $\gamma=10$.
Figure~\ref{model1}(a) represents several snapshots of interface patterns at different times.
The circular shaped pattern develops to a star shaped pattern.
The pattern is qualitatively similar to that observed in our experiments as shown in Figs.~\ref{magnetic} and \ref{timestar}.
Figure \ref{model1}(b) shows the colloidal density $n$ (solid line) along the interface (dashed line). 
The aggregated colloids are accumulated at the dent parts of the star shaped pattern owing to the sweeping effect. 
The behavior is similar to that observed in our experiments.

However, the star shaped pattern decays slowly with time in this model. 
This is  because the sweeping effect works only for the moving interface and the flat interface becomes stable owing to the effective line tension in the final stationary state in Eqs.~(\ref{model1-1}) and (\ref{model1-2}). 
This is not consistent with our experimental result.
Thus, we consider another model in which $\alpha(n)=1-n$ instead of $\alpha(n)=1$.
In this model, the diffusion coefficient in Eq.~(\ref{model1-1}) is assumed to decrease with $n$ due to the hardening effect of the interface 
when taking colloidal aggregation into account and it becomes 0 when $n=1$.
Figure~\ref{model2} shows numerical results of Eqs.~(\ref{model1-1}) and (\ref{model1-2}) with $\alpha(n)=1-n$ at $a=0.01$, $c=1$, $D_z=D_n=0.04$, $\beta=0.02$, and $\gamma=10$.
As shown in Fig.~\ref{model2}(a), the star shaped pattern is maintained as a stable pattern owing to the effect of $\alpha(n)=1-n$. 
Furthermore, we confirmed that the density $n$
of the colloidal aggregate is larger at the dent parts of the star shaped pattern than at the protruded parts as shown in Fig.~\ref{model2}(b).
This result is consistent with the experimental result. 

\begin{figure*}[tbp]
	\begin{center}
		\includegraphics[width=15cm]{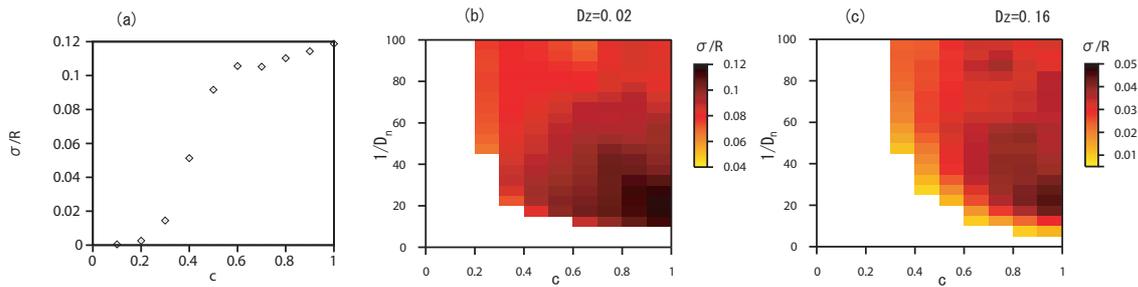}
	\end{center}
	\caption{(a) Relationship between $c$ and $\sigma/R$ in our phenomenological model in Eqs.~(\ref{model1-1}) and (\ref{model1-2}) with $\alpha(n)=1-n$ at $1/D_n=15$, and $D_z=0.02$.
		(b) Color plot of $\sigma/R$ in a parameter space of  $c$ and $1/D_n$ for $D_z=0.02$. Only the parameter region of $\sigma/R>0.04$ is plotted. 
		(c) Color plot of $\sigma/R$ in a parameter space of $c$ and $1/D_n$ for $D_z=0.16$.  Only the parameter region of $\sigma/R>0.005$ is plotted.}
	\label{modelphase}
\end{figure*}

We investigated $\sigma/R$ for the patterns obtained from Eqs.~(\ref{model1-1}) and (\ref{model1-2}) with $\alpha(n)=1-n$ for various values of $c$ and $D_n$. 
Here $R$ and $\sigma$ are respectively the average value and the standard deviation of the distance $r(\theta)$ from the center of mass to the interface. 
Parameters $c$ and $1/D_n$ qualitatively correspond to $c$ and $\mu$ in our experiments,
and we could compare our theoretical results with our experimental results in Fig.~\ref{phase}. Figure~\ref{modelphase}(a) shows a relationship between $c$ and $\sigma/R$ at $1/D_n=15$ and $D_z=0.02$. 
Since $\sigma/R$ increases rapidly near $\sigma/R\sim 0.04$,
we investigated a phase diagram for $D_z=0.02$, 
in which only the parameters satisfying $\sigma/R>0.04$ are plotted (Fig.~\ref{modelphase}(b)). 
Moreover, Fig.~\ref{modelphase}(c) is a phase diagram for a different value of $D_z=0.16$, 
in which only the parameters satisfying $\sigma/R>0.005$ are plotted,
as $\sigma/R$ increases rapidly at $\sigma/R\sim 0.005$.
As $D_z$ is larger, the interface deformation is suppressed, 
and a therethold value of $\sigma/R$ becomes smaller. 
However, the star shaped patterns appear when the concentration $c$ is large, 
and the amplitude of the star shaped pattern takes the maximum value near $c=1$ and $1/D_n\sim 20$ in both cases of $D_z=0.02$ and 0.16. 
The numerical results shown in Figs.~\ref{modelphase}(b) and (c) are qualitatively close to those in our experiments, in that the star shaped patterns emerge at large $c$.
This numerical result of our phenomenological model suggests 
(1) the spreading of an acidic solution 
with a lower surface tension on the glycerol solution,
(2) 
colloidal aggregation of milk protein colloids caused by the denaturation 
that occurs when mixed with an acidic solution,
and (3) the accumulations of the colloidal 
aggregates toward the dent regions of the moving interface by the sweeping effect are important for the formation of the star shaped pattern.

\section{Summary}
We discovered that a star shaped pattern was formed owing to an interfacial instability 
when an acidic solution with a low surface tension 
spread on the surface of a glycerol solution including milk proteins.
In the spreading process, 
aggregates of milk protein colloids included in the glycerol solution at the interface 
between two solutions 
develop by 
the denaturation of mill protein colloids that occurs when mixed with an acidic solution included in the droplet solution, 
which leads to the formation of the star shaped pattern.
We proposed a phenomenological model that included following points;
(1) the spreading of the acidic solution on the glycerol solution,
(2) colloidal aggregation of milk protein colloids,
(3) the accumulations of the colloidal aggregates toward the dent regions of the moving interface by a sweeping effect.
The model provided the star shaped pattern which was similar to the experimental one.
Moreover, the model obtained a phase diagram of the pattern formation against a concentration of 2-methoxyethanol in the acidic solution and the viscosity of glycerol solution, 
which were control parameters in our experiments.
The phase diagram is close to that obtained from our experiments.
We discussed the formation of the star shaped pattern through a comparison between our experimental results and those obtained from our phenomenological model.  
Although the quantitative comparison was difficult, we were successful in the qualitative comparison.
One of the reasons for the difficulty is considered to be that we could not decide the values of many parameters in our model, such as $\beta$,
$D_n$, $D_z$, and $\gamma$.
As a point of the future investigation, 
it would be interesting to measure $n$ at each time and to investigate a distribution of $n$ around the pattern.
The studies would lead to developments in the physical understanding of interfacial instability in non-equilibrium systems.

\acknowledgments
We would like to thank Prof. M. Tokita, Prof. H. Honjo and Prof. S. Ohta in Kyushu University, 
Prof. T. Takami in Oita University, 
T. Komiya, A. Toshimitsu, K. Tanaka in Fukuoka Institute of technology for their fruitful discussions and suggestions. 
MS would like to thank S. Howkins in Fukuoka Indtitute of Technology for a proofreading of this paper.
This work is supported by JSPS KAKENHI Grant Nos. 15K17723 and 18K03462.


\begin{thebibliography}{0}
\bibitem{food} Mixing food color into the solution of the droplet has almost no effect on the pattern formation as the amount of the food color is negligible.
\end{thebibliography}
\end{document}